\documentclass{aa}

\usepackage{graphics,epsfig}

\newbox\grsign \setbox\grsign=\hbox{$>$}
\newdimen\grdimen \grdimen=\ht\grsign
\newbox\laxbox \newbox\gaxbox
\setbox\gaxbox=\hbox{\raise.5ex\hbox{$>$}\llap
     {\lower.5ex\hbox{$\sim$}}}\ht1=\grdimen\dp1=0pt
\setbox\laxbox=\hbox{\raise.5ex\hbox{$<$}\llap
     {\lower.5ex\hbox{$\sim$}}}\ht2=\grdimen\dp2=0pt
\def\gax{\mathrel{\copy\gaxbox}}
\def\lax{\mathrel{\copy\laxbox}}

\def\rxj{RXJ0720.4$-$3125}

\outer\def\gtae {$\buildrel {\lower3pt\hbox{$>$}} \over 
{\lower2pt\hbox{$\sim$}} $}
\outer\def\ltae {$\buildrel {\lower3pt\hbox{$<$}} \over 
{\lower2pt\hbox{$\sim$}} $}

\begin{document}

\title{
First XMM-Newton observations of an isolated neutron star:
RXJ0720.4$-$3125
\thanks{Based on observations
obtained with XMM-Newton, an ESA science mission with instruments
and contributions directly funded by ESA Member States and the
USA (NASA).}
}

\author{
Frits Paerels \inst{1}, Kaya Mori \inst{1}, 
Christian Motch \inst{2},
Frank Haberl \inst{3},
Vyacheslav E. Zavlin \inst{3},
Silvia Zane \inst{4},\\
Gavin Ramsay \inst{4}, 
Mark Cropper \inst{4},
Bert Brinkman \inst{5}
}

\institute{
Columbia Astrophysics Laboratory, Columbia University,
550 West 120th St., New York, NY 10027, USA
\and 
Observatoire Astronomique, CNRS UMR 7550, 11 Rue de l'Universit\'e,
F-67000 Strasbourg, France
\and
Max Planck Institut f\"ur Extraterrestrische Physik,
Giessenbachstrasse, D-85748 Garching, Germany
\and
Mullard Space Science Laboratory, University College London, Holmbury
St. Mary, Dorking, \\ Surrey, RH5 6NT, UK
\and
SRON Laboratory for Space Research, Sorbonnelaan 2, 
3584 CA Utrecht, the Netherlands
}

\authorrunning{
F. Paerels et al.}

\titlerunning{
X-ray Spectroscopy of \rxj
}

\date{Received 2 October 2000/ Accepted 26 October 2000}

\abstract{
We present the high resolution spectrum of the isolated neutron star 
\rxj, obtained with the Reflection Grating Spectrometer on {\it
XMM-Newton}, complemented with the broad band spectrum observed with
the EPIC PN camera. The spectrum appears smooth, with no evidence for
strong photospheric absorption or emission features. We briefly
discuss the implications of our failure to detect structure in the
spectrum.
\keywords{stars: atmospheres --
	  stars: individual (\rxj) --
	  stars: neutron -- X-rays: stars
	  }
}

\maketitle

\section{Introduction}

There is now compelling evidence that the {\it ROSAT} source 
RXJ0720.4-3125, first
discovered in a systematic survey of the galactic plane (Haberl et al.
1997), is a nearby isolated neutron star. 
Its spectrum appears very soft ($kT_{\rm e} \sim 80$ eV), and there
is little attenuation of the X-ray flux by interstellar absorption. 
There is a plausible optical counterpart 
(Motch \& Haberl 1998; Kulkarni \& van Kerkwijk 1998),
and the measured X-ray to optical flux
ratio basically rules out an accreting (low-mass) X-ray binary, or an
isolated white dwarf. Perhaps the strongest piece of evidence is
the detection of a fairly shallow, quasi-sinusoidal modulation of the
X-ray flux, at a period of 8.39 sec (Haberl et al. 1997). The
stability of this period, as measured with 
{\it ROSAT} over the period 1993-1996, suggests that we
are seeing modulation by the spin of a compact object, most likely
a neutron star.

The sinusoidal
shape of the modulation in turn suggests that it is due to changing
visibility, and possibly limb-darkening effects, of a hot polar cap
(or polar caps),
against a cooler atmosphere. Such a region could be heated by
accretion from the interstellar medium, with the accretion stream
focused by a magnetic field of order $B_{\rm surface} \lax 10^{10}$ G
(the field can be no stronger, or else it would prevent accretion onto
the surface, given the spin period and likely values for the density
of the insterstellar medium). Alternatively, the object could be 
releasing heat from decay of a very strong magnetic field, which would
at the same time modulate the heat flow through the atmosphere due to
the influence of the field on the thermal conduction in the
direction transverse to the field (Greenstein \& Hartke 1983). 


These issues can be directly addressed by studying both
the spin history, as well as the detailed
photospheric emission spectrum of the object.
With {\it XMM-Newton}, we have recently obtained data of superior
sensitivity that will allow us to pursue both these possibilities.

In this {\it Letter} we present a preliminary analysis of the
high-resolution spectrum of \rxj, obtained with the Reflection Grating
Spectrometer (RGS; den Herder et al. 2001), which covers the range
5-35 \AA\ at an approximately constant
resolution $\Delta\lambda = 0.05$ \AA. Data were obtained
simultaneously with the EPIC focal plane imaging cameras (Turner et
al. 2001), of which the PN camera (Str\"uder et al. 2001)
offers the combination of high
sensitivity and high time resolution. This latter dataset will be used
to investigate the evolution of the spin period. Unfortunately,
however, we currently lack the satellite dynamical data to perform an
accurate barycentric correction to the photon arrival times, and we
defer an accurate measurement of the spin period to a forthcoming
paper. 

\section{Photospheric spectroscopy}

\subsection{Data analysis}

{\it XMM-Newton} (Jansen et al. 2001)
observed our target \rxj\ on 13 May 2000, during the
Calibration/Performance Verification phase of the mission, for a total
of 62.5 ksec (continous viewing).
The data were processed using the standard RGS pipeline software,
provided within the SAS (Scientific Analysis System) package. 
Reconstructed events are placed on a common dispersion-angle and
CCD pulse height scale.  Bad pixels are removed from the data, and an
exposure map is calculated to correct for the resulting spatial
modulation of the sensitivity. Based on the known geometry of the 
spectrometers, dispersion angles are converted to wavelength (for
details of the preflight and in-flight wavelength calibration, we
refer to den Herder at al. [2001]). 
The resulting photon list is four dimensional: each event is labeled
with wavelength (or dispersion angle), cross-dispersion coordinate,
CCD pulse height, and CCD frame number (time). A spectrum is assembled
by imposing a joint spatial/CCD-spectral filter on the data. This
filter also effectively selects between spectral orders, which appear
offset in dispersion angle/CCD pulse height space. 
Background is determined from repeating this extraction for a region
on the focal plane detector parallel to the source spectrum, but
offset in the cross-dispersion direction. The response of the
spectrometer associated with the choice of spatial/pulse height
filters can be generated automatically from an explicit model for the
spectrometer (optics and detectors).

Given the 8.39 sec periodicity of the source, we tried to optimize
time resolution and sensitivity of the RGS, in order to perform
phase-resolved spectroscopy. In principle, the RGS focal plane CCD
detectors can be read out in the cross-dispersion direction
in continuous clocking mode. This provides a photon arrival time
resolution of $\approx 15$ msec, but at the expense of the
cross-dispersion imaging information. This limits the application of
this High Time Resolution mode to bright sources only. 
On the other hand, 
the standard readout sequence of the nine CCD chips that make up one
focal plane detector requires 5.7 sec, which is too long to allow for
a meaningful phase resolution in the case of \rxj. We therefore
decided to operate one spectrometer (RGS1) in standard imaging mode
(5.7 sec time resolution), while we chose a custom readout sequence 
for the other spectrometer. We interleaved a readout of CCD chip nr 3
in RGS2 with the sequential readouts of the other chips. This chip
covers the approximate wavelength range 24-28 \AA, which is therefore
read out every 1.1 sec. This 
compromise between time resolution and spectral coverage was based on
the fact that the spectrum appears as a very soft ($kT_{\rm BB}
\approx 80$ eV) blackbody in the {\it ROSAT} PSPC, which we 
expected to peak around 20 \AA\ in the RGS. In principle, therefore,
we can phase resolve the 24-28 \AA\ region of the spectrum. In
this paper, however, we will restrict the discussion to the
phase-averaged RGS spectrum.

\begin{figure}
\resizebox{\hsize}{!}{\includegraphics{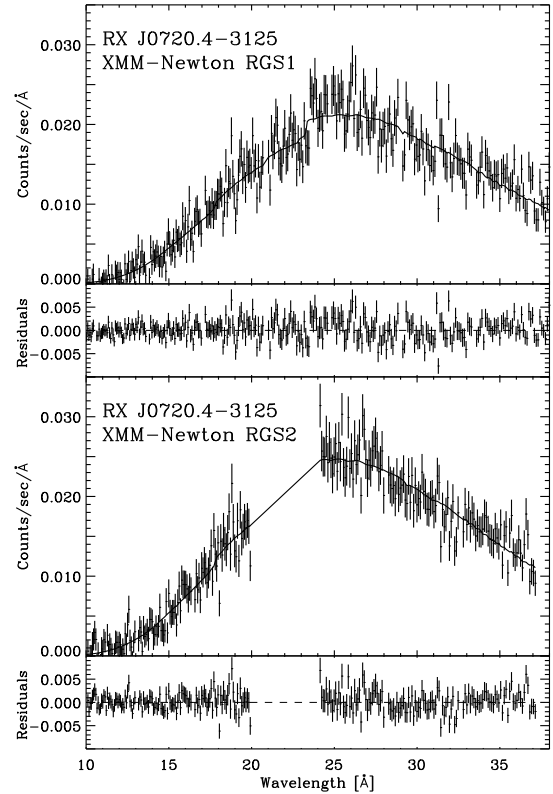}}
\caption{
The spectrum of \rxj, as observed with both RGS1 and 2. The data have
been binned in 0.1 \AA\ bins. The range $20.1-23.9$
\AA\ is not present in the data from RGS2, because of malfunction
of the drive electronics for one CCD chip in this spectrometer. 
Average total source count rate is 0.3 counts s$^{-1}$ in first order,
per spectrometer.
Superimposed is a blackbody spectral shape (see text). The
slight modulation of the RGS1 post-fit residuals
around $\sim 23$ \AA\ is due
to the presence of an O K edge in the RGS efficiency which has not yet
been fully calibrated.
}
\end{figure}

\subsection{Blackbody fits}

Figure 1 shows the spectra from the two RGS's. Overlaid on the data is
a simple blackbody spectrum, which, somewhat suprisingly and
disappointingly, turns out to be an acceptable fit to the data.
The best fitting parameters for a joint RGS1/2 fit are 
$kT_{\rm BB} = 84.2 \pm 0.4$ eV, and a column density of 
$N_{\rm H} = 2.7 \pm 0.3 \times 10^{20}$ cm$^{-2}$.
At first sight, one might be tempted to identify coherent features in
the post-fit residuals. But careful inspection shows that none of
these appear significantly in both spectrometers simultaneously, and
so we conclude that the spectrum appears smooth, to within the limits
implied by the counting statistics fluctuations in this observation.
Based on our experience with similarly smooth continuum spectra in the
RGS, we roughly estimate that features with an equivalent width
larger than
approximately one-fifth of the spectral resolution (or $W_{\lambda}
\gax 0.01$ \AA) should have been detectable.

The spectrum obtained with the PN camera, although of lower spectral
resolution than the RGS spectrum, covers a wider band than RGS, and
has higher signal to noise per resolution element, and we
use it to investigate possible subtle broad-band deviations from the
simple black-body shape.
A PN spectrum was accumulated using single-pixel events for which
a relatively advanced response matrix exists. From the total exposure
of 62.5 ks the last $\sim$10 ks and some shorter intervals could not be
used for the analysis due to strong background flaring which saturated
the data handling system. The total average source count rate is
6.1 counts s$^{-1}$ (not yet corrected for deadtime).
The spectrum was satisfactory fit with an
absorbed blackbody model. The best fit parameters are a column density
of 
$N_{\rm H} = 5.8 \pm 0.4 \times 10^{19}$ cm$^{-2}$,
and a temperature of $kT_{\rm BB} = 86.0 \pm 0.3$ eV with formal 
1$\sigma$ errors. 
The column density is a factor of two lower than 
derived from the {\it ROSAT} PSPC spectrum by Haberl et al. (1997), which
is probably caused by calibration uncertainties at the very low energies
in both instruments.
There is also a formal discrepancy between the column densities
measured with RGS and PN, probably due to the fact that RGS is less
sensitive to small column densities, and the presence of a small
remaining calibration uncertainty around the O K edge in RGS. Fixing the
column density at the best fitting PN value would slightly increase
the RGS blackbody temperature, to $kT_{\rm BB} = 89.2 \pm 0.4$ eV.

A combined fit of the RGS, EPIC-PN, and {\it ROSAT} PSPC spectra 
is shown in Figure 2. Both RGS spectra are superimposed, and all
spectra have been rebinned such that each spectral bin contains at
least 20 counts.
The best fit column density and temperature are  
$N_{\rm H} = 6.0 \pm 0.4 \times 10^{19}$ cm$^{-2}$ and 
$kT_{\rm BB} = 
86.2 \pm 0.3$ eV, and these values are dominated by the PN 
spectrum due to its high statistical quality.
The absolute normalizations
are still preliminary as accurate exposure and deadtime calculations
are not finalized yet. Residuals in the PN fit around 0.28 keV are caused
by inaccurate filter transmission data near the carbon K-edge of the
thin filter.

\subsection{Neutron star atmospheres}

We also applied neutron star atmosphere models to the RGS and
PN data. The X-ray spectra emitted by strongly magnetized
atmospheres should exhibit a number of spectral features. First
of all, there are electron and proton cyclotron
resonances, at energies of $E_{\rm B_{\rm e}}=11.6~B_{12}$~keV and
$E_{\rm B_{\rm p}}
=6.3~B_{12}$~eV, respectively, where $B=B_{12}~10^{12}$~G is
the neutron star surface magnetic field (cf.~Pavlov et al.~1995; Zane
et al. 2000).
Absence of any significant absorption features in the RGS spectra in the
$0.35-1.25$~keV range would appear to exclude magnetic fields of
$B\simeq (0.3-2.0)\times 10^{11}$ and $(0.5-2.0)\times 10^{14}$~G.
Similarly, we do not detect a significant cyclotron emission feature,
such as has been predicted by Nelson et al. (1993),
for a Coulomb heated neutron star atmosphere. We note that these
conclusions are relatively robust, because cyclotron features are
expected to appear in the spectrum regardless of the surface
composition, assuming that the atmosphere is not predominantly neutral
(which may be the case for very strong fields). 

As far as atomic transitions are concerned, their energies and shapes
are of course strongly dependent on the conditions in the atmosphere,
and our non-detection of any strong spectral features can therefore
only place model-dependent constraints on the stellar parameters.
Roughly speaking, there are four possible classes of atmospheric
models: those with, and without a strong magnetic field, and those of
pure hydrogen composition as opposed to metal-enriched or even pure
metal composition. We regard any predictions with respect to either
of these parameters as sufficiently uncertain that it makes sense to
compare our spectrum to predictions for all
four cases.

\begin{figure}
\resizebox{\hsize}{!}
{\includegraphics[angle=-90,clip,bb=100 40 555 715]
{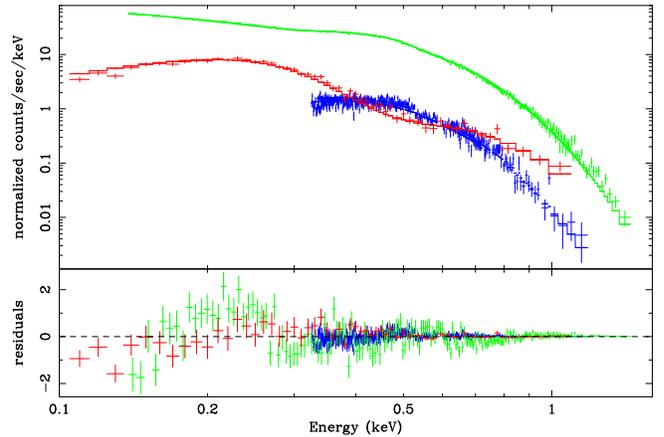}}
\caption{
Combined blackbody fit to the EPIC-PN (green), RGS (blue) and ROSAT
PSPC (red) spectra of \rxj.
}
\end{figure}

We start with pure hydrogen atmospheres.
At temperatures below $10^6$~K, the spectra of magnetized hydrogen
atmospheres have pronounced features in the soft X-ray range 
due to bound-bound transitions in the hydrogen atom
(Pavlov \& Zavlin 2001).
The energy of the strongest transition is estimated by the formula
$E\simeq 0.052 + 0.009~\ln~B_{12} + 0.006~B_{12}$~keV.
This excludes magnetic fields of $B>4\times 10^{13}$~G.
On the other hand, 
fits with hydrogen atmosphere models
for $B<10^{10}$~G and $B=10^{11}-10^{13}$~G (which have no discrete
structure in our band) require a
surface temperature of $\sim 3\times 10^5$~K,
significantly lower than that given by the blackbody fit.
However, the shapes of these very low or high B-field
atmospheric spectra are sufficiently similar to
blackbodies (of very different effective
temperature, however) that 
we cannot distinguish between blackbody
and hydrogen atmosphere models on statistical grounds. 

Low-field models with significant 
metal abundance (either solar abundances, or pure iron)
do not fit the data because of the strong 
atomic absorption features in the
spectra (Rajagopal \& Romani 1996; Zavlin et al.~1996). 
The case of strongly magnetized metal-enriched atmospheres is more
difficult to assess due to the relative scarcity of relevant
calculations. The strongly magnetized pure Fe spectra presented by 
Rajagopal et al. (1997) show significant absorption
structure in the RGS band (for sufficiently high $T_{\rm eff}$ that the
radiation is detectable), but the structure may be subtle enough 
(dense, and small contrast) that
it may not be easy to detect in data of finite resolution
and statistical quality. We regard the absence of strong features in
our spectrum as inconclusive in this respect. 

\section{The X-ray pulsation}

We have accumulated counts from the EPIC-PN detector over the energy ranges
0.1--0.4, 0.4--0.8 and 0.8--1.2 keV and phase-averaged them over 42 phase
intervals on the 8.391 sec ephemeris of Haberl et al. (1997). We
show the intensity curve in the 0.1--1.2 keV energy band together with the
hardness ratio (soft/medium) in Figure 3. Because of its poorer time
resolution we have not used the EPIC-MOS data.

The general shape of the phased intensity curve is similar to that 
found by Haberl et
al. (1997) although the uncertainties are smaller. It appears
approximately symmetrical and sinusoidal; the amplitude is $\sim15$\%. Very
interestingly, the hardness ratio is also seen to vary, in the sense that it is
softest around flux maximum, but the amplitude is smaller ($\sim10$\%).
The
phasing of the hardness ratio curve is slightly but significantly earlier than
the intensity curve. A cross-correlation of the two curves indicates that the
phase difference  between maximum intensity and maximum hardness
is $\Delta\phi = -0.048$.

\begin{figure}

\psfig{file=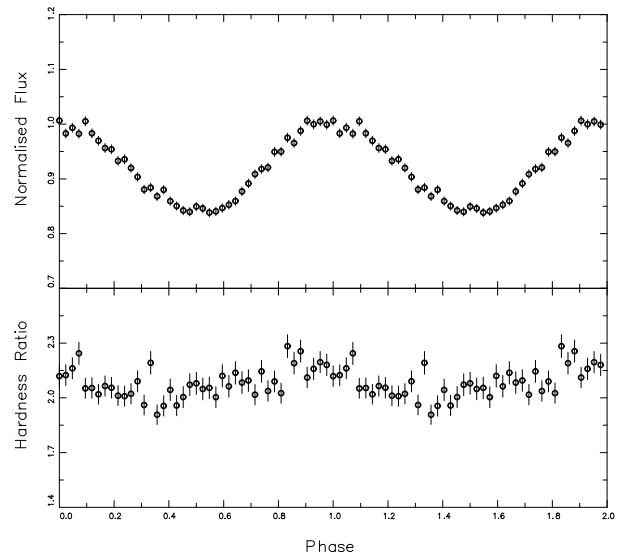,width=8 cm}

\caption{The total flux (top) from EPIC-PN phase 
folded on the 8.391 sec 
period of Haberl et al. (1997) with the (0.1 to 0.4)/(0.4 to
0.8) keV hardness ratio (bottom). The flux is normalised at phase 0.0 }
\label{fig:lc} 
\end{figure}

Because of the uncertainty in absolute timing for our data, we are
currently unable
to refine the ephemeris to the level required to discern any changes in
the period of the X-ray pulsation with respect to the period as
measured with {\it ROSAT}. A Fourier analysis of the data using an
arbitrary starting time yields only one significant period, at 8.3912 sec,
consistent with that reported by Haberl et al.  (1997).

The high S/N ratio of the X-ray pulsation in these data provides the
opportunity for a more detailed modeling  of the angular and spatial
distribution of the radiation than was possible from the {\it
ROSAT} data, and this analysis will be presented in a companion paper
(Cropper et al. 2001).

\section{Conclusions}

We present the atmospheric spectrum of the isolated neutron star
\rxj, as measured with the RGS and EPIC-PN instruments on 
{\it XMM-Newton}. We find that the spectral shape is smooth, without
evidence for strong spectral features. The absence of 
electron or proton cyclotron resonances in the RGS band would
appear to exclude
average surface magnetic field strengths of 
$B\simeq (0.3-2.0)\times 10^{11}$ and $(0.5-2.0)\times 10^{14}$~G.
The implications of the apparent absence of atomic transitions are
strongly model dependent, and we can therefore not rule out either
a pure hydrogen atmosphere or a metal-rich atmosphere, if the magnetic
field strength satisfies certain restrictions.

\begin{acknowledgements}
The Columbia group is supported by the US National Aeronautics and 
Space Administration.
We gratefully acknowledge comments from the referee, George Pavlov.
\end{acknowledgements}

\end{document}